\documentclass[sigconf]{acmart}

\newcommand {\changes}[1]{\textcolor{black}{#1}\normalfont}
\usepackage{xspace}
\usepackage{array}

\newcommand{\markC}{Mark\xspace}
\newcommand{\skylar}{S1\xspace}
\newcommand{\me}{ME102\xspace}
\newcommand{\stanford}{Stanford} 

\newcommand{\shelbyD}{Shelby\xspace}
\newcommand{\lauren}{S2\xspace}
\newcommand{\arch}{Arch438X\xspace}
\newcommand{\iowa}{Iowa State\xspace}

\newcommand{\paulE}{Paul\xspace}
\newcommand{\desma}{DESMA160-4\xspace}
\newcommand{\form}{DESMA22\xspace}
\newcommand{\ucla}{UCLA}

\newcommand{\james}{James\xspace}
\newcommand{\raw}{R.A.W.\xspace}
\newcommand{\jessica}{S4\xspace}
\newcommand{\princeton}{Princeton}

\newcommand{\ben}{Ben\xspace}
\newcommand{\itp}{ITP-Subtraction\xspace}
\newcommand{\tangible}{ITP-Tangible Interaction\xspace}
\newcommand{\nyu}{NYU}

\newcommand{\george}{George\xspace}
\newcommand{\vivek}{Vivek\xspace}
\newcommand{\adam}{Adam\xspace}
\newcommand{\desinv}{DESINV190/290-9\xspace}
\newcommand{\berkeley}{Berkeley}

\newcommand{\nadya}{Nadya\xspace}
\newcommand{\gabrielle}{Gabrielle\xspace}
\newcommand{\hcde}{HCDE598\xspace}
\newcommand{\uw}{University of Washington}

\newcommand{\jennifer}{Jennifer\xspace}
\newcommand{\sam}{Samuelle\xspace}
\newcommand{\mat}{MAT594X\xspace}
\newcommand{\ucsb}{University of California Santa Barbara}
\AtBeginDocument{%
  \providecommand\BibTeX{{%
    \normalfont B\kern-0.5em{\scshape i\kern-0.25em b}\kern-0.8em\TeX}}}

\copyrightyear{2021} 
\acmYear{2021} 
\setcopyright{rightsretained} 
\acmConference[CHI '21]{CHI Conference on Human Factors in Computing Systems}{May 8--13, 2021}{Yokohama, Japan}
\acmBooktitle{CHI Conference on Human Factors in Computing Systems (CHI '21), May 8--13, 2021, Yokohama, Japan}\acmDOI{10.1145/3411764.3445450}
\acmISBN{978-1-4503-8096-6/21/05}

\begin{document}

\title[Remote Learners, Home Makers]{Remote Learners, Home Makers: How Digital Fabrication Was Taught Online During a Pandemic}


\author{Gabrielle Benabdallah}
\affiliation{%
 \institution{University of Washington}
 }
\author{Samuelle Bourgault}
\affiliation{%
 \institution{University of California, Santa Barbara}
 }
 \author{Nadya Peek}
\affiliation{%
 \institution{University of Washington}
 }
\author{Jennifer Jacobs}
\affiliation{%
 \institution{University of California, Santa Barbara}
 }

\renewcommand{\shortauthors}{Benabdallah, et al.}

\begin{abstract}
Digital fabrication courses that relied on physical makerspaces were severely disrupted by COVID-19. As universities shut down in Spring 2020, instructors developed new models for digital fabrication at a distance. Through interviews with faculty and students and examination of course materials, we recount the experiences of eight remote digital fabrication courses. We found that learning with hobbyist equipment and online social networks could emulate using industrial equipment in shared workshops. Furthermore, at-home digital fabrication offered unique learning opportunities including more iteration, machine tuning, and maintenance. These opportunities depended on new forms of labor and varied based on student living situations. Our findings have implications for remote and in-person digital fabrication instruction. They indicate how access to tools was important, but not as critical as providing opportunities for iteration; they show how remote fabrication exacerbated student inequities; and they suggest strategies for evaluating trade-offs in remote fabrication models with respect to learning objectives.

\end{abstract}



\maketitle

\section{Introduction}
The COVID-19 pandemic created drastic societal changes at a global scale.
In the United States, a public health emergency was declared in early March 2020. In response to stay-at-home orders and social-distancing restrictions, higher education pivoted to online instruction. This change posed challenges for all types of learning. Educators had to adopt new forms of remote instruction with limited time to plan or share approaches. Classes that were centered around physical making required particularly radical changes because universities were forced to shut down physical workshops, labs, and studios. Educators across art, design, and engineering had to rapidly develop new strategies to compensate for the loss of these spaces.

In this paper, we examine the impacts of remote instruction for a particular form of physical making: digital fabrication. Physical making offers unique learning opportunities~\cite{Martinez_Stager_2016}. Digital fabrication extends these opportunities by enabling students to design and manufacture custom physical objects through a combination of computer-aided design and machining (CAD and CAM) and physical computer-numerical-control (CNC) machines~\cite{10.1145/3335055.3335070}. Digital fabrication technologies are often a central component of  makerspaces---shared workshops with access to tools and materials that support physical making. Makerspaces are increasingly prevalent in universities~\cite{rosenbaum2017dragons}, providing students with access to shared digital fabrication tools and software such as 3D printers, laser cutters, and CAD/CAM software, as well as opportunities for community support through fostered cultures of making and tinkering \cite{martin2015promise}.

Despite losing access to digital fabrication equipment and in-person communities, many educators still held their digital fabrication classes in the Spring of 2020~\cite{jacobsinteractions}. Given the unique challenges of teaching digital fabrication without a makerspace, we sought to understand what happened during those classes. Our work is guided by two research questions. \textbf{First, how did people teach digital fabrication remotely during the pandemic?}
In particular, we wanted to examine how instructors remotely taught computer-aided design and computer-controlled fabrication, how they provided and organized community, and what trade-offs they had to consider in the process.
\textbf{Second, how can we learn from instructors' efforts to teach digital fabrication in a crisis to improve remote instruction of digital fabrication in the future?} The pivot to remote instruction was the result of a terrible crisis; however, it also created a unique opportunity to examine new strategies for learning through digital fabrication. We sought to understand what elements of these strategies were effective and how they could be improved in the future.

As digital fabrication researchers, as well as educators and students who taught or took remote digital fabrication courses in the spring of 2020, the authors of this paper were both observers and subjects of the phenomena we examined. As a result, our research is structured around analysis of remote fabrication instruction in both our own courses and in the courses of others. We used a preliminary analysis of our course outcomes to guide a formal set of interviews with instructors and students in six remote fabrication courses from different universities. These interviews examined peoples' experiences planning, teaching, and participating in remote digital fabrication courses, as well as the challenges and opportunities that emerged from the remote format.

Our paper makes the following contributions. First, drawing from both our classes and the classes of others, we define and document five models of remote fabrication instruction that were used over the spring of 2020. Second, through a recounting of our course outcomes and a thematic analysis of our interviews, we surface themes on shifts in labor caused by remote fabrication access, learning opportunities of remote fabrication instruction, approaches to gaining tacit knowledge remotely, and remote collaborative practices for physical making. These themes highlight assumptions about home-work, and remote-work, as well as the tensions that arise from different ways of combining them. Third, 
we discuss what was lost and what was gained in the remote format, what was crucial about work performed by instructors and students, and what factors contributed to equity in outcomes. Combined, these contributions have implications for human-computer interaction (HCI) researchers studying digital fabrication and learning. Moreover, as the risks of the novel coronavirus persist and the future of in-person instruction remains uncertain, our work provides practical details on viable approaches for remote instruction for physical making in the future.

\section{Background}





Digital fabrication encompasses a wide range of practices. At a high level, it can describe any form of computer-controlled fabrication. This means digital fabrication contains many different elements, including computer-aided design (CAD), robotic path planning and computer-aided manufacturing (CAM), computer-numerically controlled (CNC) processes, and (robotic) placement or assembly. These processes happens across length scales, ranging from the fabrication and assembly of Frank Gehry architecture \cite{coleman2017any} to the nanometer scale fabrication of micro-electromechanical systems such as the accelerometers in a game controller \cite{walker1996focused}. HCI contributes to digital fabrication research in many ways, including through novel tools for computational design \cite{10.1145/3025453.3025927, schmidt_design--fabricate_2013}, materials \cite{10.1145/3242587.3242625, metamaterial}, fabrication \cite{10.1145/3173574.3173723,10.1145/2984511.2984553}, and collaboration \cite{10.1145/142750.142767, 10.1145/3313831.3376621}.

\changes{Teaching digital fabrication might take place in anywhere from a cleanroom \cite{moore2004laser}, to a mechanical engineering shop \cite{10.1115/DETC2006-99723}, to an architecture studio \cite{eversmann2017digital}. University digital fabrication research labs may have equipment that rivals industrial digital fabrication production factories, featuring large scale 6-axis robotic arms, milling machines, water jet cutters, and other pieces of \$100k+ equipment \cite{icd, taubman, princeton}. The courses we studied were slated to be taught in university spaces that ranged in tool sophistication and application from large robot arms to programmable embroidery machines. }

\changes{Beyond differences in equipment, courses incorporating digital fabrication can also differ in their learning goals. 
While some courses focus on developing particular skills such as designing in 3D or fabricating with a CNC mill \cite{whalen23augmenting}, others might emphasize more abstract learning goals, such as providing students with the environment in which they will conduct self-directed projects while managing resources such as materials, shared equipment, and time \cite{wilczynski2017value}, or using making for critical inquiry \cite{nieusma2016making}.
Managing spaces with diverging goals has unique challenges, including equipment cost, staffing, hours of operation, rent, community organization, maintenance, and safety.}

The rise of the maker movement \cite{dougherty2016free} and increased demand for makerspaces (academic and otherwise) has led to the development of maker-oriented, lower-cost digital fabrication equipment. These more affordable machines have increased access to digital fabrication tools and reduced cost of managing spaces with digital fabrication capabilities. The growth of makerspaces has also led to research on the efficacy of makerspaces as a learning environment \cite{10.1145/2559206.2579405}. Early advocates of makerspaces in formal education include Mike and Ann Eisenberg, who argued that hands-on interacting with materials can offer a tangible way of thinking through important and expressive ideas \cite{eisenberg1998middle}, and Paulo Blikstein, who stated that digital fabrication and maker culture could be considered the ``the ultimate construction kit'' with significant advantages for interdisciplinary and contextualized learning, powerful experiences, and team building \cite{blikstein2013digital}. 
Scholarship in digital fabrication and learning is now extensive and covered in new places including the FabLearn conference \cite{fablearn}, first held in 2011, which focuses on hands-on learning and the role of digital fabrication in education, and International Symposium of Academic Makerspaces (ISAM) \cite{isam}, first held in 2016, which focuses on starting and running academic makerspaces. 

\changes{Makerspaces make up more than just tools in a space. They are also places for gathering, peer-learning environments, and an attitude \cite{martin2015promise}. The makerspace environment shapes the ways students learn, therefore integrating makerspaces into formal education has not been without growing pains. Researchers have found that social interaction and discourse, especially as means to build community and maker attitudes, are crucial for learning in K-12 \cite{sorryteacher} and other \cite{martin2015promise} makerspaces. 
Makerspaces located at universities increasingly show a diversity of implementation, from large digital fabrication research labs to small student groups focused on making.
We refer to all spaces where digital fabrication was taught on campus as makerspaces, despite their breadth. Regardless, none of the courses we surveyed were able to work on campus. Our study is unique, as it was held at a time of unprecedented changes to higher education.}

\changes{Well before the pandemic, websites such as Instructables, Thingiverse, and YouTube provided online community gathering spaces for making and sharing designs. These online spaces have their own online-specific challenges in terms of onboarding newcomers, welcoming diversity, and encouraging sharing and remixing \cite{10.1145/3121113.3121214, 10.1145/2957276.2957301, 10.1145/2702123.2702175}. Nonetheless, online maker sites demonstrate a thriving practice of online documentation, sharing experiences, and encouraging participation. Many of the instructors we surveyed drew from such sites when restructuring their courses.}

HCI contributes crucial analysis of the promises and practices of maker culture by engaging with and unpacking the complex social, cultural, and economic conditions that makers operate within \cite{10.1145/2556288.2557132, 10.1145/2699742}. These critical analyses improve the culture and spaces in which we teach and learn, and we aim for the work in this paper to contribute to this discussion.

\section{Methods}
Our research is centered on two datasets: 1) autobiographical from our own remote fabrication courses from the spring of 2020, and 2) interviews with instructors and students who taught or attended remote fabrication courses at other universities. In this section we outline our methodology for assembling and analyzing these two datasets to provide context to the claims we make in following sections. To contextualize our methods and analysis, we also provide background on the research team.

\subsection{Author Background}
\nadya and \jennifer are professors at public universities who collaborate in research on digital fabrication. They both taught remote graduate-level digital fabrication courses in the spring 2020. \gabrielle and \sam are PhD students in interdisciplinary art, design, and engineering departments who research design and making. \gabrielle and \sam were students in \nadya and \jennifer's courses, respectively. 

\subsection{Preliminary Analysis of Author Courses}
Following the conclusion of the Spring 2020 academic quarter, \nadya and \jennifer theorized that deeper examination of approaches to remote digital fabrication could inform instruction efforts in the future. They initiated their research efforts by analyzing the outcomes of their own courses. They collected public online posting of student projects and written student reflections. They met regularly for three weeks to review this data and discuss their experiences as instructors. They extracted preliminary themes from their course data through the collaborative writing and editing of a written reflection. Their writing process was organized around 1) examining of the effects of the remote format to learning outcomes and 2) evaluating of the impacts of at-home fabrication equipment~\cite{jacobsinteractions}. 

\subsection{Interview and Analysis Methods for External Instructors}
Following the analysis of \nadya and \jennifer's courses, \sam and \gabrielle were brought on as collaborators. Together, we used the preliminary themes from \nadya and \jennifer's course analysis to determine selection criteria and interview structure for instructors and students in remote fabrication courses at other universities.
We identified potential interview candidates through a short online survey that collected information on the general approaches university educators used to teach digital fabrication remotely. 
We received 23 survey responses over a period of one week. We selected eight individuals representing six different courses for interviews---five instructors via the survey and three additional co-instructors of the same course who were recommended by a colleague. 
We selected instructors who represented a range of models of remote fabrication instruction to study how different people compensated for the loss of in-person makerspaces. 
Instructors were our primary focus, however we also conducted interviews with three students in three of the external courses to contrast instructor and student experiences. 

Interviews were conducted remotely over video conference and lasted one hour. Interviews with instructors and students focused on their experiences planning, teaching, and participating in remote digital fabrication courses, challenges and opportunities that emerged from the remote format, and how the experience impacted their perspective on teaching or participating in digital fabrication in the future. 
All interviews were audio recorded and transcribed. To analyze the data we conducted a reflexive thematic analysis~\cite{braun_thematic,braun2019reflecting} focusing on latent themes. Following each interview, the authors met and discussed initial aspects of the data. After all interviews were complete, each author open-coded a subset of interview transcripts. 
\gabrielle performed an initial conceptualization of the codes into preliminary themes and all authors discussed these initial themes. Based on the outcomes of this discussion, \jennifer performed a secondary review and refinement of the themes identified by \gabrielle. The themes were further refined in a final group discussion. Out of a list of eleven themes, we selected a subset of four that we believed were the most important due to their consistent presence in all the interviews and the amount of data we compiled on them. 

\subsection{Limitations}
We relied, in part, on autobiographical data. The shutdown offered a unique opportunity to study the uncommon practice of remote digital fabrication instruction in its early stages. We incorporated autobiographical data in this research because we used an instruction model that was not present in our external data. Furthermore, by including an analysis of our own experiences, we provide context for the motivation of this research and the conclusions we made.
%
We compared courses in different departments and subjects; however, instructors apply digital fabrication technologies for different learning objectives. This factor was evident in our data and impacted the approaches individual instructors took when selecting models for remote instruction. We saw value in surveying the ways remote digital fabrication supports learning across domains, however future studies which examine remote fabrication in a specific area will likely provide domain-specific insights.
%
\changes{We discussed how the remote learning format exacerbated uneven access to resources for students. We believed this was a point of particular importance for current and future remote digital fabrication instruction. Our data did not allow us to provide a more detailed picture of how discrepancies among students' living situations impacted their learning during the pandemic. Further research is needed to understand and address this key factor in successful and equitable remote teaching of digital fabrication.}

\changes{\section{Course Summaries}}
\begin{figure*}
  \centering
  \includegraphics[width=0.85\textwidth]{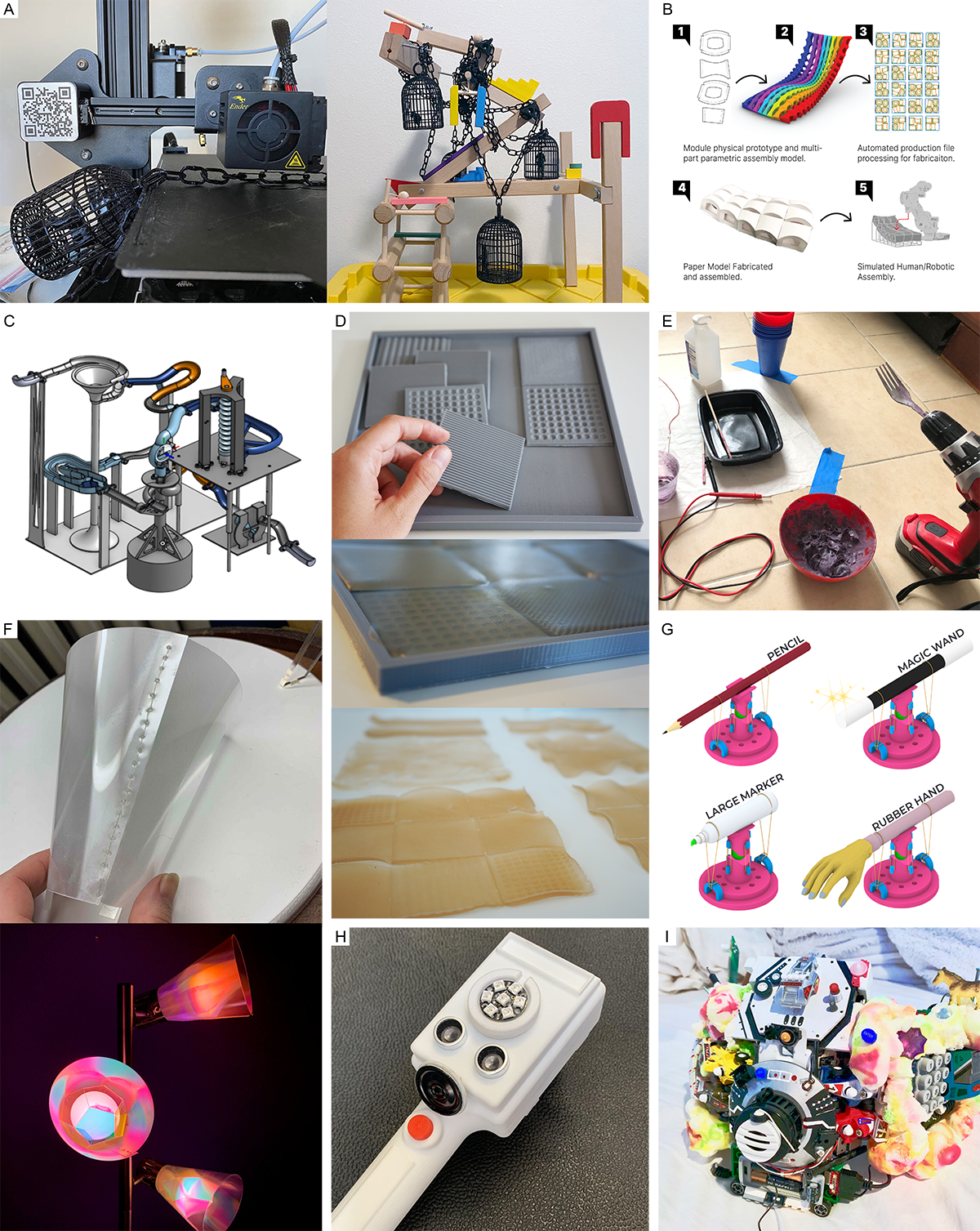}
   \caption{A wide range of student work was produced in remote digital fabrication classes in Spring 2020. A) Yanrong Chen’s sculpture with many interlocking parts iteratively printed in \hcde B) Design of a space frame from a single node to robotic assembly in \raw C) A marble maze collaborative CAD project in \me D) \sam's bioplastic cast in 3D printed molds in \mat E) Conductive silicone mixed with a fork-drill in kitchen containers by Pippa Kelmenson in \tangible F) Vinyl lamp iterations by Aidan Lincoln in \itp G) Pen holder designs for a robotic arm by Samuel Rushenberg in \arch H) Jaideep Cherukuri, Scout Handford, Jahangir Abbas Mohammed, Abrar Syed \& Miyuki Weldon combining electronics and 3D prints in \desinv I) Kat Sung using found and recycled objects in \desma.} 
  \Description{Figure 1: 9 fabrication projects including physical objects and CAD simulations organized in a grid.
Figure 1A: On the left half, a 3D printer in the process of printing the chain part of a 3D printed birdcage with a black filament. On the right, three similar 3D printed birdcages hanging from a small wooden structure.
Figure 1B: A visualization of 5 steps to create an architectural space frame including 1) the design of a single node, 2) the CAD simulation of the space frame, 3) the generation of two-dimensional production file for fabrication, 4) the paper model fabricated and assembled and 5) the simulated human/robotic assembly.
Figure 1C: A complex CAD figure representing a multi-part marble run structure.
Figure 1D: On the top third, a hand holding a 3D printed texture square designed to fit in a square mold. On the middle third, the bottom of the mold is covered with texture squares and cast with bioplastic. On the bottom third, 8 parts of dry bioplastic made with this mold and aligned on a table next to each other.
Figure 1E: A bowl with a purple substance of non-cured conductive silicone in it next to a take-out container, chopsticks, plastic cups, a fork in a drill that served to mix the silicone on an apartment floor.
Figure 1F: On the top half, hand holding a semi-transparent vinyl cone attached on one side with a thread. On the bottom, a lamp made of three bulbs with colorful vinyl cones used as lampshades.
Figure 1G: 4 CAD figures of the same pencil holder with different objects in it to attach to a robot arm: the first one holds a pencil, the second a magic wand, the third a large marker, and the last one a rubber hand.
Figure 1H: A white 3D printed device with visible distance sensors, a red button, a small speaker, and electronics, that can be fixed on a white cane to help visually impaired people to detect obstacles.
Figure 1I: A mask with a cyberpunk aesthetic made of found and recycled objects and colored construction foam}
  \label{fig:figure1}
\end{figure*}

In total, we analyzed the outcomes of eight remote courses involving digital fabrication ( table~\ref{tab:table1}). In this section, we summarize the structure of each course, focusing on the models instructors used to retain access to digital fabrication technologies and hands on making.

\subsection{Author Course Summaries}
\nadya and \jennifer's courses used the same model for remote digital fabrication instruction: students were shipped hobbyist 3D printers and all digital fabrication instruction was oriented around these machines (\textbf{at home machines}).

\subsubsection{\hcde~- Digital Fabrication}
\hcde was a course developed by Nadya in Human-Centered Design and Engineering, an interdisciplinary department at the University of Washington. Twenty students enrolled in this quarter-long course in Spring 2020, supported by two TAs. The course introduced students to CAD and prototyping tools for making physical artifacts. For remote instruction students were asked to purchase a \$250 3D printer alongside hand tools (e.g., calipers and Exacto knives) and materials (e.g., 3D printing filament, silicone, plaster, and cardboard.) The total cost per student was {$\pm$}\$350. 

\subsubsection{\mat~- Computational Fabrication}
\mat was a course developed by Jennifer in Media Arts and Technology, an interdisciplinary graduate department at the University of California, Santa Barbara. 
The course included twelve students from Media Arts and Technology 
and Computer Science. The course emphasized computational fabrication; students used programming languages to design for and control digital fabrication machines. For the Spring 2020 quarter, \jennifer used a combination of research funds and departmental resources to purchase low-cost 3D printers, PLA filaments and additional supplies, such as specialty filament, casting materials, and electronic and lighting components, to send to students. The total cost per student ranged from \$250-350. 

\subsection{External Course Summaries}
 We identified four additional models for remote digital fabrication across the six external courses we surveyed: simulation of fabrication with CAD/CAM (\textbf{simulation}),  ordering from online fabrication vendors (\textbf{online-vendors}), converting the university makerspace to a service (\textbf{makerspace-to-jobshop}) , and having students or instructors fabricate parts for other students with at-home equipment (\textbf{instructor/student-technicians}).
 In addition to these models of digital fabrication access, we observed three supplemental strategies for retaining hands on making: shipping materials and hand tools directly to students (\textbf{material shipping}), requiring students to independently source their own materials and hand tools (\textbf{student sourcing}), and having students rely on materials and tools already in their homes (\textbf{home materials}).

\setlength{\extrarowheight}{.5em}
\begin{table*}[t]
  \centering
  \begin{tabular}%
  {>{\raggedright\arraybackslash}p{.14\textwidth}%
  >{\raggedright\arraybackslash}p{.12\textwidth}%
  >{\raggedright\arraybackslash}p{.06\textwidth}%
  >{\raggedright\arraybackslash}p{.2\textwidth}%
  >{\raggedright\arraybackslash}p{.17\textwidth}%
  >{\raggedright\arraybackslash}p{.19\textwidth}%
  }
    & \multicolumn{2}{c}{\small{\textbf{Interview Subjects}}}  & & & \\
    \cmidrule(r){2-3}
    {\small\textit{Course}}
    & {\small \textit{Instructors}}
    & {\small \textit{Students}}
    & {\small \textit{Fabrication Access Models}}
      & {\small \textit{Field}} 
      & {\small \textit{School}}  \\
    \midrule
    \hcde & \nadya Peek & N/A & at-home machines, material shipping & HCI/Engineering/ Design & \uw \\
    \mat & \jennifer Jacobs & N/A & at-home machines & HCI/CS/New Media Art & \ucsb \\
    \me & \markC Cutkosky & \skylar& simulation, online-vendor, student sourcing, home materials & Engineering & \stanford   \\
    \arch & \shelbyD Doyle & \lauren & simulation, makerspace-to-jobshop & Architecture & \iowa  \\
    \desma & \paulE Esposito & N/A & instructor-technician, material shipping & Fine Arts & \ucla\\
    \raw & \james Coleman & \jessica & student-technician & Architecture & \princeton\\
    \itp & \ben Light & N/A & at-home machines, student-technician, student sourcing & Fine Arts & \nyu\\
    \desinv & \vivek Rao, \adam Patrick Hutz, \george Moore & N/A &  online-vendor & Engineering/Design & \berkeley\\
  \end{tabular}
  \caption{Summary of Surveyed Courses}~\label{tab:table1}
\end{table*}

\subsubsection{\me~- Foundations of Product Realization}
\me was a quarter-long Mechanical Engineering course taught by \markC Cutkosky at Stanford University. Sixty engineering undergraduate students enrolled. Approximately 10 TAs were also assigned to this course.
The course objective was to engage students with a design-to-fabrication process through the making of iterative prototypes using digital fabrication machines in a shared workshop. To adapt to remote-learning, the focus of the course shifted to emphasize online collaboration in CAD. 
Students constructed physical projects as low fidelity prototypes using materials at home (e.g., cardboard, foam core, Exacto knives, and glue.) In one assignment, instructors used on-demand fabrication services to 3D print students' designs. The total cost was less than \$100 per student and covered by the department.

\subsubsection{\arch~- Architectural Robotics}
{\arch} was developed and taught for the first time in Spring 2020 by 
\shelbyD Doyle
in the Architecture Department at Iowa State University. This semester-long course became remote mid-semester. Twenty-four undergraduate students enrolled. \arch acted as an introduction to robotics and aimed to expand students' perception of the role of robots in architecture. The course was designed to give hands-on experience in making small robots and in using a KUKA industrial robot recently acquired by the ISU Computation+Construction Lab (CCL). The CCL lab also included digital fabrication equipment, robotic devices, hand tools and power tools. In the shutdown, the robots became unavailable and the goal of the course shifted to focus on simulation and speculative design. 

\subsubsection{\desma~- Survival Tools in Weird Times}
This quarter-long course was a variation of \form~- \textit{Form}, and was created specifically for remote instruction in Spring 2020. It was taught by \paulE Esposito in the Department of Design and Media Arts (DMA) at the University of California, Los Angeles. Twenty-one undergraduate art students with different levels of experience with fabrication enrolled. To adapt to the lack of fabrication lab equipment and materials, \desma focused on the theme of survivalism and its intersection with maker culture. \paulE had two 3D printers and a sewing machine at home and offered to print and sew the designs of his students and mail them the results. Students who wanted to hand sew their own designs were also shipped a sewing kit. The course budget included \$12 kits for each student and an additional \$500 materials budget which \paulE used for 3D printing filament. 

\subsubsection{\raw~- Robotic Architecture Workshop} 
\raw was a remote workshop that was taught at Princeton University during Summer 2020 by \james Coleman through the Black Box Research Group in the Architecture department. The workshop was two weeks long and \james and two graduate TAs met with students six times over this period. Six students enrolled, including civil engineering Ph.D. students, architecture graduate students, and undergraduate students from the Engineering and Architecture departments. The objective was to familiarize participants with the design-to-fabrication-to-assembly workflow required to make space frames using sheet metal. The in-person format would have involved making metal parts in an industrial shop then assembling them robotically. This experience was replaced with robotic simulation and paper prototyping using a Silhouette Cameo 4 vinyl cutter. Some of the students received a \$280 vinyl cutter and all students had a materials stipend of \$90. 

\subsubsection{\itp \& \tangible}

\itp \& \tangible were two semester-long graduate classes taught and co-taught by \ben Light at the New York University Tisch School of the Arts within the Interactive Telecommunications Program (ITP). Fifteen students enrolled in \itp and fourteen enrolled in \tangible.
\itp was intended to be an introduction to subtractive fabrication techniques with hands-on experience with machines and \tangible was intended to focus on making physical interfaces. For in person courses, each student paid \$300 in lab fees and could either buy their own material or use lab scrap for free. Both courses moved online half way through the semester. To adapt, \ben shipped a Silhouette Cameo vinyl cutter to each \itp student and focused on two-dimensional fabrication techniques for the rest of the course. The machines cost \$200 each and were covered by the department. 
\ben removed the digital fabrication aspect of \tangible during the second half of the semester to focus mainly on physical computing. 


\subsubsection{\desinv~- Technology Design Foundations}

\hfill\\
\desinv was taught by \vivek Rao, \adam Patrick Hutz, and \george Moore within the Jacobs Institute for Design Innovation in the College of Engineering at the University of California, Berkeley. 
The course was originally designed for graduate students but most of the twenty students enrolled during the Spring 2020 semester were undergraduates. \desinv was developed to familiarize students with a human-centered design process. This process included sketching ideas, conducting interviews and analyzing data in order to validate a design, prototyping at different levels of fidelity, using digital fabrication machines, and integrating interactive digital systems to fabricate objects. When the university makerspace closed midway through the curriculum, the instructors decided to use on-demand fabrication services for the rest of the semester. The students received a budget of \$250 per team to order parts from several online fabrication vendors. 

 

\section{Remote Instruction With At Home 3D Printers}
\changes{In this section, we describe the themes that emerged from analysis of author-led courses, \hcde and \mat, in response to our first research question (How did people teach digital fabrication remotely during the pandemic?). We focused on: 1) the impacts of at-home 3D printers on student workflows and domestic activities, 2) the unique learning opportunities of hobbyist machines in comparison to workshop equipment, and 3) the ways students developed tacit knowledge while engaged in remote instruction.}


\subsection{Impacts of At-Home 3D Printers}

Shipping printers to students' homes created a situation where students were simultaneously living with printers and creating objects for personal use with them. Several students also took personal initiative or assignment contexts to use the printers to design home goods or to repair or augment existing objects in their home. One \mat student created a program that generated designs for a customizable self-watering planter, and one student in \hcde created a modular lamp integrated with internal lighting components to create different patterns of light diffusion. 

Projects such as these conformed to many aspects of personal fabrication; the objects were custom-designed and fabricated by their maker as opposed to mass-manufactured and purchased. At-home access to the machines did not simplify or accelerate production, or lead to fundamentally new design and manufacturing workflows. Instead, producing these products required students to engage in design workflows that reflected elements of real-world design, manufacturing, and craft. Students in both \hcde and \mat engaged in learning, design, testing, and iteration; and required peer support when fabricating personal objects for home use. These processes were odds with product-focused visions of personal fabrication where consumers create custom objects with minimal effort and knowledge. 

The presence of the printers in students’ homes also resulted in changes to students' routines and daily activities. Because students often lived with roommates or occupied small studio apartments, they often kept their printers in their bedrooms. This, coupled with long print times and heat, smells, and machine noises generated by printers, resulted in students coordinating their schedules around their printers. These factors also created additional stress when prints failed.
The students managed to accommodate the requirements of the printer, but it was not difficult to envision scenarios where such constraints would be infeasible.
There were also elements of at-home 3D printing that provided important forms of stress relief and pleasure. Students in both courses repeatedly expressed their delight at being able to make physical objects and seeing the products made by their classmates. 

\subsection{Learning Opportunities of Hobbyist 3D Printers}
The use of at-home printers had unique opportunities when compared with how students accessed machines in a workshop. Unlike staff-managed workshop equipment, individual printers required students to learn about machine maintenance. The Ender 3 Pro required assembly and fine-tuning the printer could greatly improve printing outcomes. \nadya built this opportunity directly into her curriculum by making the printer’s assembly and initial calibration one of the first assignments in \hcde. By the end of the spring quarter, students in both courses had tuned and modified their machines to a degree that went significantly beyond the manufacturer documentation. Several students in \hcde upgraded components (such as the fans or power supplies) or 3D printed components to improve performance (such as clips for wire management, holders for work surface illumination, or filament guides). These activities enabled students to familiarize themselves with the machine's implementation details and performance possibilities in a form that would not have been feasible in a shared-use setting.

The at-home setup also allowed for constant access to the printers, which in turn allowed students to iterate extensively on their designs. Repeated design iterations were common in both courses and went beyond simple optimizations. For example, one student in \mat went through multiple iterations to find a successful printing strategy for sculptures generated from complex photogrammetry data that she had previously only used for digital designs. 
Students were also able to create different kinds of artifacts by developing custom fabrication processes for their machines. In some cases this involved close integration of manual manipulation and machine fabrication. One student in \hcde created a complex sculpture of interlocking chains and birdhouses, which were printed as interlocking structures by pausing the printer at key moments and inserting previously-completed parts (shown in Figure \ref{fig:figure1}A). Completing the sculpture involved many tens of hours of print time that were interspersed with regular adjustments or actions made by the student.


The quality and sophistication of many student projects in both classes suggested that at-home 3D printers provided unique learning opportunities for machine maintenance and modification, while supporting increased design iteration. Such opportunities are often obstructed when machines are shared and maintained by others.

\subsection{Gaining Tacit Knowledge Remotely}
The use of at-home 3D printers enabled students to work across CAD, CAM, and CNC throughout the courses. We observed students iterating in CAD based on initial machine prints, learning to modify CAM settings based on model geometry, and iterating on machine and material settings based on settings they looked up and tuned. 
These outcomes demonstrate how learning opportunities in integrating CAD, CAM, and machine operation remained present in the remote format with some key differences. Students were limited to learning these concepts with one form of additive fabrication. In a makerspace they would have had the opportunity to learn CAD-CAM-CNC design processes for additional subtractive fabrication processes. This limitation was highlighted in a subtractive CNC/CAM assignment in \mat where several students ran into errors of scale---attempting to fabricate parts that were much too large or small for the target (simulated machine) or similarly selecting tooling that was much too small. Direct exposure to subtractive milling hardware and tooling would have likely provided a way to inform this process in a way that was less feasible through simulation. 

All digital fabrication machines place constraints on what can be fabricated. Producing successful products requires learning how to design for these constraints in CAD, how to engage in incremental testing when working with new equipment and materials, and how to systematically adjust machine and CAM parameters to optimize for different geometries. The hobbyist printers imposed more severe constraints than machines we had used in prior courses, but they still enabled students to develop these forms of knowledge in the domain of additive manufacturing.

\subsection{Summary}
\changes{The combined outcomes from two author-led classes that used the at-home machine model suggested that remote instruction with distributed hobbyist 3D printers is a viable method for graduate-level digital fabrication courses. Shifting the workshop to the home led to complex forms of personal fabrication while creating a mix of positive and negative lifestyle changes. This model offered learning opportunities that were less feasible in shared makerspaces,  such as maintenance and increased iteration. It also enabled the understanding of tacit knowledge associated with the constraints of the 3D printers.}

\section{Remote Instruction With Other Fabrication Models}
\changes{In this section we describe the themes that emerged from our analysis of external remote fabrication courses. 
We conceptualized themes across three dimensions that built on the analysis of the author-led courses. 1) We further examine how home life was impacted by remote fabrication by highlighting how other models of instruction introduced new forms of at-home labor for instructors and students. 2) We examine how simulation, online-vendor, and makerspace-to-jobshop models of machine access shaped learning outcomes in comparison to the at-home machine model of our courses. 3) We contrast the ways instructors in different disciplines valued tacit knowledge and attempted to preserve it in a remote learning environment. We also explore a fourth theme (4) unique to the external data. We compare strategies for collaboration in remote fabrication courses through the experience of student teams.}



\subsection{New Home Labor Through Remote Fabrication Access}\label{fabaccess} 
Instructors in the majority of the courses we surveyed worked hard to create some form of remote digital fabrication access. Each model of access created new forms of labor for instructors.
In cases where instructors and students fabricated parts for other students with machines in their homes, they took on the role of shop technicians. In \raw, one student and two TAs acted as vinyl cutter operators for the other five participants. In \itp several students took initiative to create their own job shop. As \ben described:
\begin{quote}
\textit{One person bought an Othermill or a Bantam mill and someone bought 3D printer. [The students] were all sort of like, ``I've got this. If you need a part, I'll run one.''} 
\end{quote}

When students or instructors worked as fabrication technicians, they took on non-trivial tasks of monitoring production and delivering parts. \markC raised the concern of relying on TAs as technicians rather than educators.
\begin{quote}
    \textit{I need to be a little careful\dots TAs didn't sign up to become a printing service. They signed up to become teaching assistants and that's what they want to do.}
\end{quote} 
\james described how \raw fabricators were not able to mail the parts in time and resorted to photographing the pieces, assembling them and sending the photos to the students. 
These delivery issues were similar to the challenges instructors encountered when using professional online fabrication services. \me and \desinv used online fabrication vendors, and in both cases students experienced delays in receiving the parts. The logistics of using online vendors disrupted students' ability to personally test and revise their parts. \markC described how \me TAs tested on-demand printed parts for students in the lab and \adam in \desinv explained how shipping delays constrained students to ``only one iteration on the timeline.''

Different models of fabrication access resulted in different degrees of use, depending on how they were implemented. \arch had the option of using the university makerspace as a jobshop, however students could only receive their parts by picking them up directly. \lauren pointed out that the makerspace-as-jobshop model was ``really only an option for a few people,'' adding that ``a lot of people that I know moved home, which could be a couple hours away\dots a couple of states away.'' The student/instructor-technician model also resulted in limited use in the courses we surveyed.  
For \desma, only two of the 21 students had their parts printed by \paulE. He described how the students who used his home-printing service were those most motivated to develop their existing CAD and digital fabrication skills. 

In comparison to the makerspace-as-jobshop and instructor/student-technician models, there was evidence that the at-home machines model led to higher rates of machine access and use. \ben described how students who received machines were able to use them at greater rates, and at irregular hours:
\begin{quote}
    \textit{The thing I loved about the vinyl cutters more than anything that actually came out of it, is that students got to live with the machines. And I think that's really the only way to get good at it, right?\dots You get a crazy idea and then you immediately make it. \dots you somehow learn [the machine] inside and out. It starts having quirks that you know. \dots That's something that never happened in the past because no one had the machines. They did it here [on campus] and then they left and there were 10 people behind them waiting for their turn. I think it just may not have been the machine they wanted, but having total access to something and the time
    \dots being trapped indoors with nothing but your vinyl cutter... you know, they learned it.}
\end{quote}
When considering the limited use of the instructor/student-technician model in comparison to at-home machines it is important to note that instructors made using this model optional. If it were required, the use rates would likely have been different.

Similar to student experiences in our courses, the presence of machines at home was also disruptive to home life for instructors and students to a certain extent. For example, \shelbyD described noise interference from her 3D printers when she was on Zoom calls. She could also hear the printers at night when going to bed. Machines were not the only source of at-home disruptions. The expectation to do any physical prototyping could also be a burden for some students. \skylar in \me described being unable to prototype effectively in his home, saying ``There's not really a lot of places in my house where I can do that kind of work.''

Overall, instructors relied on a wide range of strategies to palliate the absence of traditional fabrication spaces. Whether they chose at-home fabrication, a student/instructor-technician model or a an online-vendor model, the choices they made were closely tied to the learning objectives of their course. No models were clearly superior or inferior; rather, each emphasized different aspects of digital fabrication practice and each surfaced new forms of labor.

\subsection{Learning Opportunities of Remote Fabrication Instruction}\label{learnings} 
Remote instruction required instructors to make major changes to curricula in a short period. Similar to the experience of the authors, these changes created new learning opportunities, which were often the result of how instructors responded to the constraints of their chosen model of fabrication access. 

\markC altered \me to focus on collaborative CAD with minimal elements of hands-on making.~\paulE created an entirely new course (\desma) because his department determined it was infeasible to teach the original digital fabrication course in a remote format with limited preparation time. Creating a new course gave \paulE freedom to experiment with new forms of hands-on making including manual sewing and knot tying.

Instructors also changed how they interacted with students. Four instructors said they increased the amount of pedagogical support they provided to individual students. \paulE had weekly progress check-ins with each of his 23 students, ranging from five to twenty minutes. \markC and his TAs swapped longer lectures for more targeted sessions so that the students could ``have more detailed coaching on the projects they're working on.'' \markC's student, \skylar, felt this form of coaching was very effective in comparison to his experience in some in-person Mechanical Engineering courses.

The online-vendor and student-technician models created conditions where only some students had access to machines or physical parts. Instructors found they could use this structure to better simulate the multi-party design workflows of industry. As \james described:
\begin{quote}
\textit{I think it's artificial to say that the designer is the fabricator, and is also the erector, [and] is also the project manager. And so I think there's actually something interesting about the fact that we were forced to be separate. That made it easier to show the tensions between these groups. As opposed to me simulating that in a workshop environment where I would separate teams into different groups to force the sort of miscommunications that typically happen.}
\end{quote}

The teaching staff of \desinv also found that the online-vendor model aligned better with some students' learning objectives. \adam described how some students were more interested in learning how to fabricate and prototype on their own whereas other students are more interested in the design process and ``don't really care about the actual product.'' 

While learning opportunities in machine use were reduced in classes that relied on simulation, online-vendors and student/instructor-technicians, instructors created new opportunities in response to these constraints. Because these models reflected the realities of distributed expertise and resources in industrial design and manufacturing, they offered the chance for students to learn about supply chains and division of labor. It's important to note that exposing these new opportunities required substantial additional instructor and TA labor.

\subsection{Gaining Tacit Knowledge Remotely}\label{tacit}
In all the surveyed courses, instructors shared the perception that physical making was a critical component of the learning objectives. Describing the ethos of his department,
\ben mentioned that the ``first ugly cardboard prototype'' is ``like a rite of passage, I think for every student.'' He added that learning CAD is only one component of the fabrication pipeline and that physical making is required to understand materiality.
\begin{quote}
    \textit{I think material is something that is rarely thought of in the CAD stage and, or the CAM stage, even, other than speeds and feeds. I think they learn that not all material is equal. I think they learned that how more, you know, like be prepared for it'll work 20 times in the 21st time it won't work or, you know, that there's a reality to these things and it's not magic. I think that translates no matter what machine or whatever you're doing.}
\end{quote}

This sentiment was echoed by students and instructors in other courses. Overall interview subjects felt that CAD and simulation alone could not teach students the critical material elements of digital fabrication including fit, surface finish, and tolerance. 



\shelbyD also described the technical understanding that results from physical making. Before the pandemic, in one of her regular first assignments she required students to create a ``cast without undercuts.'' 
She described how students often did not initially grasp the concept of designing for undercuts. Only when ``they pour the plaster'' do they ``understand it.'' \shelbyD believed that this physical experimentation was important for students because it simultaneously helped them learn techniques and develop confidence when using the machines.


As universities shut down, the instructors we surveyed felt the need to emphasize the important learning factors for physical making, sometimes pushing back against detractors in the process. For \shelbyD, moving online reinforced the importance of in-person teaching of physical making, especially in a context where she was ``constantly having to kind of defend the value of that kind of teaching'' in her own institution: 
\begin{quote}
\textit{I do think moving online made it me more aware of how valuable that in-person teaching was, if that makes sense. I've never been very good at explaining it\dots
it matters that we stand in a space together and we make things and there's a sense of community and shared intelligence that comes out of that.}
\end{quote}
\shelbyD was hopeful that the shift to remote instruction would underscore the critical importance of in-person fabrication courses in the future. 
For \lauren, one of \shelbyD’s students, the complications in reviewing physical objects remotely made the full-online format difficult to adhere to. She described the awkwardness of having to showcase a physical project through video calls in comparison to walking around, touching, or otherwise interacting with such a project in an in-person studio critique. 

For courses that required only some students to engage in fabrication, like \raw, or courses where fabrication was optional, like \desma, students' motivation to purse learning elements of physical making was sometimes reduced. In the \raw, \jessica, who was already highly skilled in digital fabrication, expressed ambivalence about her role as a student-technician for the class: 
\begin{quote}\textit{It wasn't a waste of time, but it would have been easier if someone else had done it, but I still think it was useful to me to like actually do it myself, but I still feel like the other participants still learn equally.}
\end{quote}


All the instructors we interviewed stressed the importance of hands-on making to acquire the tacit knowledge required for digital fabrication. Not all curricular changes reflected this concern. Instead, instructors made decisions about hands on fabrication in relation to the specific aspects of the larger fabrication ecosystem they originally sought to target in their course. 
In cases where instructors chose to preserve tacit learning opportunities, instructors and TAs undertook additional labor in the form of acquiring and distributing equipment and materials. 

\subsection{Remote Collaboration for Physical Making}\label{collab} 
Pre-pandemic, in-person collaboration was often a central component of both professional and student digital fabrication practices. The instructors and students we interviewed worked to maintain elements of collaborative design and construction of physical objects despite being unable to meet in person. Student collaboration was built into the structure of 3 classes we surveyed. In \me, \desinv and \arch, students were assigned a team for the duration of the class. Initially, remote collaboration was demotivating for students accustomed to collaborative physical construction in makerspaces. 
\lauren in \arch described how ``asynchronous collaboration'' was frustrating when ``you're so used to liking touching things and working together.'' 
In addition to collaborative construction, students and instructors valued the peer learning, motivation and support opportunities of physical makerspace communities. As \skylar put it:
\begin{quote}\textit{
There's something really, really fun about biking across campus to the [workshop] 
late at night and seeing all the other people working on their projects, bouncing ideas off each other, asking TAs that are there for help.} 
\end{quote}

Instructors relied primarily on online communication technologies and collaborative CAD tools to retain collaborative workflows in the remote format. Students also developed new organizational strategies to coordinate at a distance. In \me, \skylar and his team 
established a workflow in order to optimize synchronous collaborative CAD development over Zoom, where they would alternate between brainstorming, prototyping, and assembling 3D models collaboratively using Onshape, and working individually on their respective parts of the design. According to \skylar, it was actually easier to meet over Zoom than in person for CAD-based issues; they could simply ``get on zoom and fix it'' quickly. When it came to manufacturing and building physical objects, remote collaboration often involved asynchronous assembly or division of labor. 
Students teams in \desinv assigned one member---usually the one with the most prior digital fabrication experience---to receive and assemble all parts from an on-demand fabrication service. A limited number of teams sent duplicate parts to other members to enhance their  understanding of the part physicality. 

Remote CAD collaboration also required divisions of labor and advanced planning. A team-based assignment in \me required each student to design a system in CAD that interfaced with their teammates’ systems to generate a continuous marble run (see Figure~\ref{fig:figure1}C). Working remotely required teams to define the spatial placement of each 3D model in relation to the others in advance and create a modular design with different components assigned to each team member.  

In addition to the frustration students experienced transitioning from in-person to remote collaboration, later issues arose with teamwork and communication.  
\markC found that creating team cohesion over online social networks was more difficult, especially if students were new to the subject or did not know each other in person. These tensions were exacerbated when team members were unable or unwilling to use the same software tools in collaborative CAD, which produced dissatisfaction among students and discrepancies in the outcome.

In spite of these tensions, one student and one instructor saw potential benefits to the logistical challenges imposed by remote collaboration. Interviewees described that remote format provided field-specific workflows. \skylar pointed out that ``a lot of what you do now with CAD is collaborative\dots So [\me] was the most perfect training for that.'' 
In \raw, \james felt that the remote setting enabled participants to select roles in line with their interests. 
\begin{quote}
  \textit{This separation of roles I think is really interesting\dots People who are interested in the digital workflow and the file prep in the parametric design jumped into that in a physical workshop. It would have been excellent to have the people who want to be the ``hands-on folks'' designing the jigs and doing the assembly.}
\end{quote}
He described further how the remote setting could simulate the division of expertise that is common in professional architecture and manufacturing practice. 
The absence of a makerspace created significant shifts in patterns of collaboration. Students were assigned explicit roles and labor was divided based on interest and expertise. The pleasurable collaboration of in-person makerspaces was absent, however some students and instructors saw alternative learning opportunities that reflected professional design and fabrication practice.

\subsection{Summary}
\changes{The fabrication access models in the six courses that we surveyed were chosen by the instructors to comply to specific course objectives. These models created different learning opportunities depending on their implementation and often created additional labor for instructors and TAs. The use of simulation, online vendors, makerspace-as-jobshop, and student/instructor-technicians reduced the amount of tacit knowledge students could gain from operating machines but still allowed students to engage in workflows, collaborative practices, and division of tasks reflecting industrial realities. 
Similar to the authors' courses courses, students with home access to machines and physical materials were able to develop greater levels of machine familiarity and physical construction experience while undergoing disruptions and new forms of labor in their daily routines.}

\section{Discussion}
\changes{The COVID-19 pandemic called attention to implicit elements of digital fabrication instruction which, as soon as they became absent or more difficult to access, required more labor to maintain: the tacit elements of physical making; the facilitation of collaboration in the classroom; and providing equal access to resources. In this section, we discuss three main takeaways from the analysis of our data: }
\changes{\begin{enumerate}
    \item The courses’ learning objectives had a great impact on which tacit elements of digital fabrication were transmitted to students. This stresses the importance of articulating course objectives and structure over access to fabrication spaces when teaching digital fabrication, especially remotely. 
    \item Proper scaffolding, providing students with opportunities for exploration and iteration, and facilitating peer collaboration yielded stronger learning outcomes, according to our data, than a focus on access to tools and materials alone.
    \item Uneven access to both material and human resources among students was exacerbated in a remote context. Clearly defining learning objectives became critical for instructors so that they could make more informed decisions about what material resources to incorporate in their curriculum and how to manage them.
\end{enumerate}}

\changes{Each of these takeaways provides insights for our second research question (how can we learn from instructors’ efforts to teach digital fabrication in a crisis to improve remote instruction of digital fabrication in the future?).} 


\subsection{What Do We Lose When We Lose the Makerspace?}

There are many definitions of what a successful digital fabrication course looks like. This reality was brought into sharp relief during the pandemic, as instructors needed to make quick decisions on what to preserve and what to change when transitioning their course online. This is in part because digital fabrication encompasses many forms of practice. There are workflows that are directly relevant to industry, such as the production of architectural elements or medical devices. There are specific workflows developed by artists for their unique work. Individuals may practice digital fabrication as a form of craft. There are many workflows which combine elements of digital fabrication alongside elements of traditional manufacturing or craft. Each of these forms of practice corresponds to distinct categories of artifacts that can be made. The shape of what is possible in turn shapes attitudes about digital fabrication.

Because of this, the tacit learning components of digital fabrication are difficult to situate. 
While all instructors agreed that these tacit components are tied to the experiential nature of digital fabrication, how this experiential component is conveyed varied widely. 
\nadya and \jennifer opted for an at-home fabrication model, where students acquired hobbyist machines and lived with them. 
The instructors we interviewed described a range of strategies, which we can divide in two main categories: at-home fabrication (\ben) and diffused fabrication, where the whole group relied on one or a few fabricators, whether they were the instructor (as was the case in \paulE's course), the TAs (in \markC's class), other students (in \james'), the makerspace staff (in \shelbyD's) or an external fabrication service (\vivek/\adam/\george, \markC).

Both types of approaches had pros and cons. \nadya and \jennifer observed that living with machines was not without challenges for their students---with issues of noise, fumes, and space management---but when properly accommodated, provided many learning opportunities for machine maintenance, modification, and design iteration. For instructors who consider a similar fabrication model, paying particular attention to how hobbyist machines fit into the students’ living context can smooth eventual frustrations and hindrances to learning.

The instructors we interviewed who chose an at-home fabrication model also reported gains and trade-offs to this approach. In \ben's class, there were issues distributing vinyl cutters to students, resulting in two students not receiving equipment at all. Students who did get access to equipment, however, gradually became used to their vinyl cutter, exploring and trying different approaches, ultimately settling for usages that suited their interests and learning goals. The fact that the students ``got to live with the machines'' meant that they not only developed a deeper knowledge of their tool but also that they could expand their fabrication practice.

In the diffused fabrication model, the fabrication process was shared between several parties and usually circulated from students (who designed the part) to technicians (either other students, a TA, the instructor or a professional service) and back to students (either in physical or virtual form). For this model, the external data showed that particular attention needed to be paid to both the course logistics---planning timelines to receive files, debug them, print parts and ship them to students---and the course scaffolding so that students could take advantage of these resources. The experience of \paulE showed that only students who were ready in terms of skills and vision took advantage of his fabrication setup. Without proper scaffolding, students were not always motivated or comfortable using the services made available to them. 
A diffused fabrication model, however, provided the opportunity to learn another type of tacit knowledge in digital fabrication, that is the ebb and flow of collaborating on larger projects, where fabricators, designers, and project managers are often separated. In this scenario, the tacit learning component was not conveyed to students through access to tools or parts but through access to a collaborative fabrication workflow. 

What do we lose, then, when we lose the makerspace? We might think that with the loss of physical fabrication spaces, the tacit learning components of digital fabrication disappear. Instead, our data shows that these tacit components resurfaced in students' homes, in collaborative processes and in virtual environments, and that these manifestations are intrinsically linked to the course's learning objectives. For instance, the experiential aspect of digital fabrication in \james's course was tied to the level of the course (the students had experience in fabrication) and its topic (architectural robotics, which often involves multi-party workflows). For \ben's class, which focused on expression, getting students access to machines so that they could explore and create was critical. 

There is not, therefore, one set of tools or materials that will guarantee successful learning of digital fabrication. Rather, different learning objectives will result in different decisions for choosing what material resources are most appropriate for a given class. These decisions are tied to the class' level, the field of study, and the students' backgrounds. 

The challenges of teaching the tacit elements of digital fabrication were exacerbated in remote formats but also presented an opportunity to better articulate them. 
Making learning objectives explicit is crucial, as well as understanding how they are tied to certain digital fabrication practices, how they lead to specific choices in material sourcing and distribution, and how they are are sometimes at odds with other curricular goals. 
This is an occasion to reconsider the locus of experiential learning in digital fabrication not in the makerspace, but in the practices each instructor facilitates.

\subsection{``What Works is to Teach a Process'': Exploration, Iteration, Contextualization}

When analyzing our data, we found that students having the ability to explore and iterate was more important for successful learning outcomes than what means of fabrication they had access to---whether it was students 3D printing on inexpensive printers at home, or sending parts out to be fabricated. Iteration happened especially when the instructors gave assignments that encouraged exploration and experimentation, as was the case in \shelbyD's class  where the students had to come up with several versions of a KUKA robotic arm end-effector. 
Creating a space for exploration and expression for students goes hand in hand with a proper contextualization of how the approaches they learn fit into a larger landscape of computation and fabrication. 
For example, \james spent a significant amount of time explaining exactly how the problems they were going to solve with paper craft corresponded to problems they would have encountered had they been using sheet metal. 
In \vivek, \adam, and \george's class, the workflow established for students via an on-demand fabrication service recreated workflows they were likely to encounter in the workplace, according to the instructors. 

Another important factor for successful learning outcomes we observed was individual or targeted support for students. Working with a small group of students and a mentor relieved some of the anxiety of being in a large class. As remote learning lingers on the horizon, increasing the role of Teaching Assistants in mentoring might prove beneficial to students, especially as it recreates the more targeted assistance that can be found in makerspaces.

The data showed no indication that some minimal amount of equipment would be sufficient to catalyze learning. Rather, we observed that learning outcomes were more strongly tied to instructors' ability to contextualize the learning environment, challenge students, and support community and iteration. This happened in each of the courses we analyzed, but with emphasis on different aspects and practices of digital fabrication.

\subsection{Inequities in Distribution of Machines, Materials, and Labor}

The pandemic is calling attention to many existing issues, among them unequal student access to both human and material resources. These inequities became particularly prevalent in the context of digital fabrication learning, which is resource-intensive. During remote instruction, access to tools and a peer community strongly depended on individual student situations. These can vary widely, with some students having ample space to accommodate tools in their living environment as well as established rapport with peers, while others faced isolation and challenging home situations. These inequalities can lead to inequities if instructors and institutions do not work to provide and facilitate equal access of human and material resources to their entire student body.  Instructors play a crucial role in how access to resources is managed. By being specific about what the learning objectives are, they can make better decisions about what material resources are needed and how best to distribute them. 

There were many ways in which access to equipment ended up being uneven. For example, not all students had space for machines in their living quarters. These students performed additional work of packing machines when not in use, then taking them out again when working. When given credits to use towards fabrication services, some students delayed the fabrication of parts in favor of more CAD revisions. This delayed the learning of tacit elements of digital fabrication such as the unintended effects of computational design decisions on production. Shared living spaces are also not immune to unfortunate accidents, such as when the roommate of one student in \vivek, \adam, and \george's class stepped on the assembled model for the course's final showcase and entirely broke it. One student reported that ``the bigger discrepancy between students is internet connection.'' Material sent out to students in countries other than the US was often more expensive to buy and ship and sometimes impossible to get to the students.

Access to human resources is as critical as access to material ones. Open and welcoming communities for peer-learning contributed extensively to positive outcomes in the remote digital fabrication classes we surveyed. In some cases, these communities pre-dated the classes and the pandemic but in others they were scaffolded during the class. Instructors established and organized online communities. \paulE reported initiating a Discord group for students to ask questions to each other. \nadya observed the evolution of her class' online community, which remained active after the course ended. Ensuring students had access to one or several people---whether the instructor, a TA, or another student---created positive learning outcomes. This is the case not only during online teaching, but required more labor to create in a distance learning format. 

Instructors also reported inequities in labor among students. For instance, \vivek explained that despite the instructors' efforts to provide a collaborative video editing platform, students still relied on the most experienced editor. While not unique to remote instruction, these inequities were exacerbated in a remote context where asynchronous collaborative processes can be difficult and where in-person accountability mechanisms are absent. More importantly, fabrication models influenced how work was shared between peers. On-demand fabrication services meant that iteration and exploration was often not possible for students, which pushed them to rely more heavily on their more experienced peers. 

Having more materials to iterate and experiment with helped students understand possibilities and trade-offs in fabrication. Overall, the classes we surveyed did not find good ways of providing students with a centralized repository for materials, access to which is nonetheless critical to experimentation. 
For \jennifer, ensuring consistent material access to her students is crucial for the next iterations of her computational fabrication course online. \nadya considers institutional support as key to managing equal distribution of resources for students. 

As we are writing this paper, our departments are communicating that remote instruction of digital fabrication courses in the Spring of 2021 is not unlikely. Instructors and institutions can work towards developing approaches to remote instruction of digital fabrication that are not provisional but cohesive and integrated into the students' living situations. What was evident from the courses we surveyed is that with greater possibilities for planning, remote instruction of digital fabrication could, if not completely address inequities in access to resources, work towards not aggravating them and even creating new opportunities for students.
For \shelbyD, teaching her studio class this Fall meant coming up with other assignments that engage her students' creativity and surface the opportunities hidden in their living spaces, such as ``conceptual robots that [the students] build at home out of things that they have. They won't necessarily need to be mechanized.'' She added:

\begin{quote}
\textit{I think if we were teaching online on purpose rather than kind of as an emergency, I could be really, I could feel more creative about it, you know? Like, it would be fun.}
\end{quote}





\section{Conclusion}

The COVID-19 pandemic has endured in the US substantially beyond the day in spring 2020 when campuses shut down. This research shows the work of students and instructors teaching and learning digital fabrication in a crisis. 
We examined how students were provided with remote access to digital fabrication, whether through at-home fabrication or diffused fabrication, and what their respective challenges were.
We identified unique learning opportunities of remote instruction of digital fabrication, including increased opportunities to iterate with at-home equipment and increased opportunities for collaboration, documentation, and engagement through remote learning technologies.
We recounted different approaches instructors took in teaching important tacit elements of digital fabrication remotely.
We found that overall, there was no minimum requirement for equipment to still learn important elements of digital fabrication. 
Rather, it was more important that instructors framed the work, established buy-in, and supported students' iteration.
Furthermore, we called attention to the ways inequities persist across education, including remote digital fabrication education, and reiterated that it is of paramount importance for instructors and institutions to work together towards more just student experiences.
We are now in a protracted crisis, or ``the new normal.'' 
While the future remains uncertain, we hope that it will hold sustainable and equitable opportunities for students to have hands-on learning experiences, even if those learning opportunities need to happen at a safe distance.




\section{Acknowledgments}
We are grateful to all of the instructors and students who shared their experiences with us, including Mark Cutkosky, Shelby Doyle, Paul Esposito, James Coleman, Ben Light, Vivek Rao, Adam Patrick Hutz, and George Moore. We also greatly appreciate the guidance and input from Madeline Gannon, Daniela Rosner, and Audrey Desjardins.  This research was funded in part by the NSF IIS Human-Centered Computing program (\#2007045) and the UCSB Academic Senate Faculty Research Grant program.

\bibliographystyle{ACM-Reference-Format}
\bibliography{remotefab}

\end{document}